\documentclass[twocolumn]{aastex62}
\usepackage{graphicx}
\usepackage{epstopdf}

\received{July 12, 2018}
\revised{July 16, 2018}
\accepted{July 25, 2018}

\submitjournal{the Astronomical Journal}

\shorttitle{Radial velocity comparison of Gaia DR2 and RAVE DR5 survey}
\shortauthors{Deepak \& Reddy, B. E. (2018)}

\begin{document}
\title{Radial velocity comparison of Gaia DR2 and RAVE DR5 survey: a systematic offset in radial velocities among a group of highly accurate radial velocity stars}
\correspondingauthor{Deepak}
\email{deepak@iiap.res.in}

\author[0000-0003-2048-9870]{Deepak}
\altaffiliation{\url{https://orcid.org/0000-0003-2048-9870}\\choudharydeepakmehla@gmail.com}
\affil{Indian Institute of Astrophysics \\ Bangalore 560034, India\\ deepak@iiap.res.in, ereddy@iiap.res.in}

\author{Bacham E. Reddy}
\affil{Indian Institute of Astrophysics \\ Bangalore 560034, India\\ deepak@iiap.res.in, ereddy@iiap.res.in}

\begin{abstract}
Here, we report comparative study of radial velocity ($\rm RV$) data of two major surveys: Gaia Data Release 2 and RAVE Data Release 5. We restricted the sample to stars with relatively accurate radial velocities ($\sigma_{\rm RV_{Gaia}} \leq$ 2 km s$^{-1}$ or $\leq$ 2\%, and $\sigma_{\rm RV_{RAVE}}\leq$ 2 km s$^{-1}$ or $\leq$ 2\%). The difference between $\rm RV_{Gaia}$ and $\rm RV_{RAVE}$ for a majority of the sample follows normal distribution with mean = 0.28 km s$^{-1}$ and $\sigma$ = 1.49 km s$^{-1}$.
However, we found a very small group of stars ($\approx 0.08\%$ of the total) for which the difference in radial velocities between the two surveys is significantly larger with an offset of $-$104.50 km s$^{-1}$ with $\sigma$ = 4.92 km s$^{-1}$. Kinematics based on  $\rm RV_{ Gaia }$ suggest that most of the group members belong to the Galactic thin disk which agrees with the group's metallicity range of $-$1.2 to $+$0.5 ~dex suggesting the offset in radial velocity is probably due to RAVE velocity data for this particular group.   
\end{abstract}

\keywords{Catalogs -- Surveys -- Techniques: radial velocities -- Stars: kinematics and dynamics -- Stars: abundances -- Galaxy: kinematics and dynamics}

\section{Introduction} \label{sec:intro}
Gaia, an European space telescope mission meant for recording accurate astrometry of more than a billion stars in the Galaxy, began its scientific observations in July 2014 \citep{GaiaMission2016A&A...595A...1G}. Since then, Gaia has been scanning the whole sky and observing all the stars within magnitude limits of  2.0 $\lesssim G \lesssim$ 20.7.
Recently released Gaia data DR2 (hereafter Gaia DR2) provides median radial velocity (over 22 months) of about 7.2~million stars \citep{BrownGaiaDr2Summary2018arXiv180409365G}. The radial velocities are determined with the Radial Velocity Spectrometer (8450 -8720\AA) having spectral resolution of R $\sim$ 11,700 (RVS, \cite{KartzGaiaDr2Rv2018arXiv180409372K}; \cite{CropperGaiaRvS2018arXiv180409369C}). The typical uncertainty in radial velocities are within 2 km s$^{-1}$.
On the other hand, Radial Velocity Experiment (RAVE) (2003-2013), is a ground based magnitude-limited survey of stars. The selection criteria and the magnitude distribution have been discussed in \cite{SteinmetzRAVEDr12006AJ....132.1645S} and in \cite{MunariRave2014AJ....148...81M}, respectively.
Spectral region (8410-8750 \AA) with effective spectral resolution, R $=\ \lambda / \Delta \lambda$ $\sim$ 7,500 was selected to cover the Ca$_{\rm II}$ triplet which is similar to Gaia's RVS (\cite{SteinmetzRAVEDr12006AJ....132.1645S}.
The fifth data release of RAVE survey (hereafter RAVE DR5) includes radial velocities of 457,588 unique stars from 520,781 spectra, which has a typical accuracy better than 2 km s$^{-1}$ \citep{Kunder2017AJ....153...75K}.\\  

Radial velocity ($\rm RV$) is a key parameter along with accurate astrometry for computing stars kinematics. 
While selecting a sample of stars from the publication of Gaia DR2 and RAVE DR5, we noticed an offset  of about $-$104 km s$^{-1}$ in radial velocities between the two surveys for a tiny group of stars ($\approx 0.08\%$ of the total stars), although the remaining radial velocities from the two surveys turns out to be in good agreement.
Our motive for this article is to highlight the existence of this tiny faulty group and its consequences. We have not attempted to provide solutions for the discrepancy or corrections, rather we looked at different possibilities that might have caused the offset in radial velocities between the two surveys.

\section{Data Sample} \label{sec:Sample}

For this study, we adopted radial velocity data directly from a catalogue of stars which resulted  from cross matching of RAVE DR5 and Gaia DR2 (rave\_DR2\_gaia\_source.csv available at {\it Rave survey website} (\url{https://www.rave-survey.org/project/}) and added RAVE DR5 table to this using RAVE\_OBS\_ID (Unique Identifier for RAVE objects, Observation Date, Fieldname, Fibernumber). We took only those stars that are common in the two surveys and for which  both the radial velocities ($\rm RV_{RAVE}$ and $\rm RV_{Gaia}$) are available. This resulted in 456,316 stars out of the total number of 512,971 stars.

\section {Analysis}
\begin{figure}[ht!]
\gridline{
			\fig{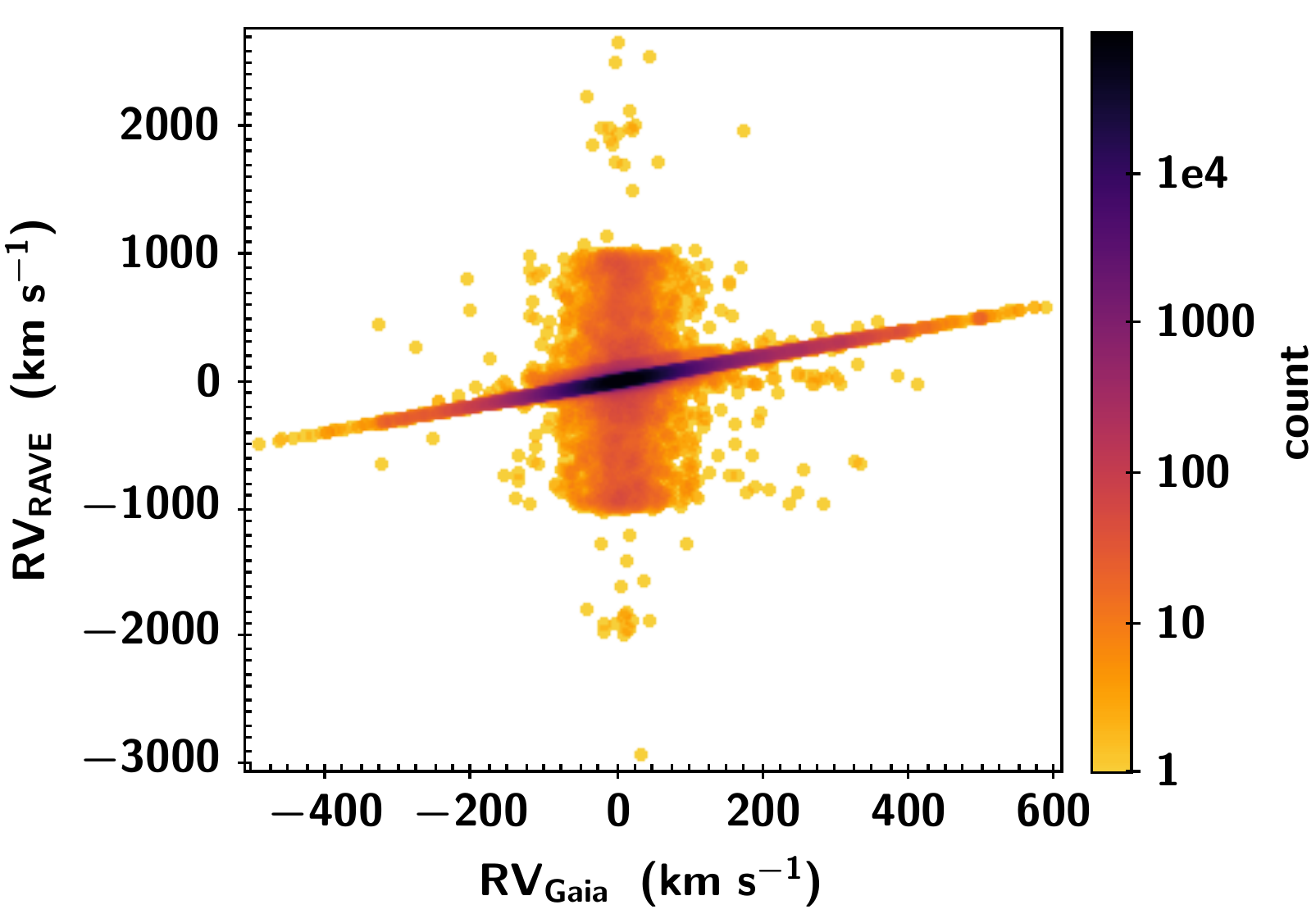}{0.25\textwidth}{(\emph{a})}
          	\fig{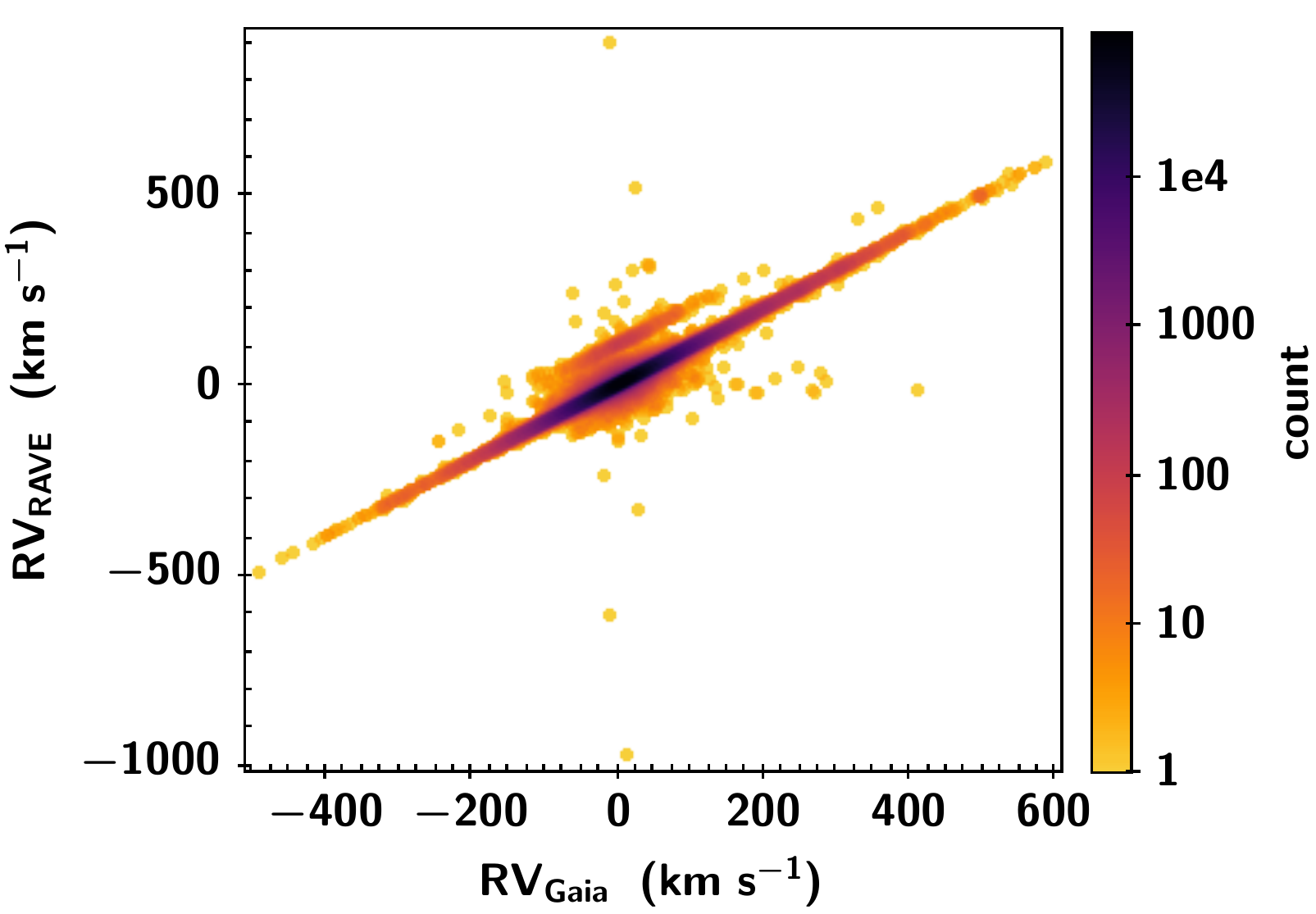}{0.25\textwidth}{(\emph{b})}
          }
\gridline{
			\fig{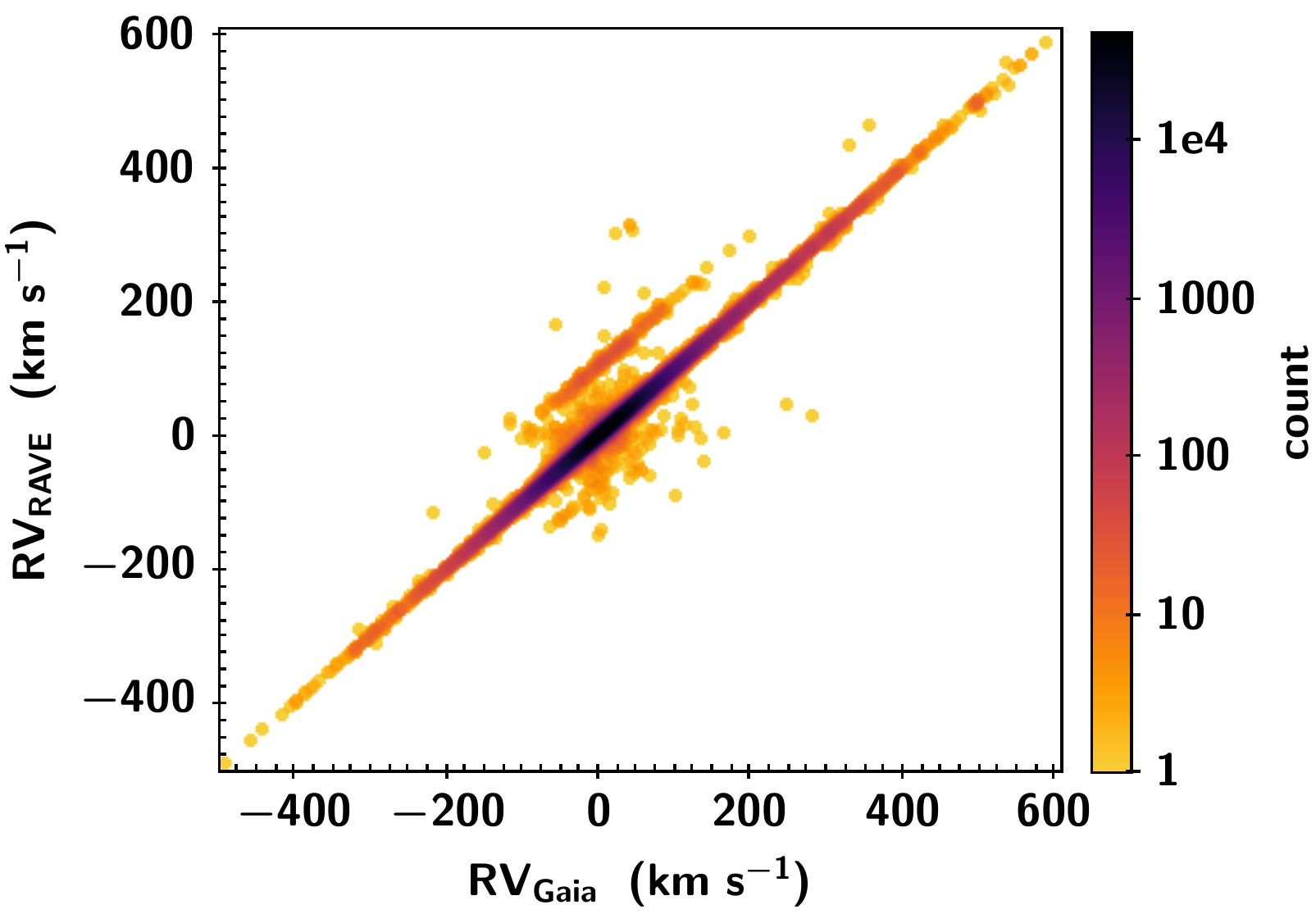}{0.25\textwidth}{(\emph{c})}
          	\fig{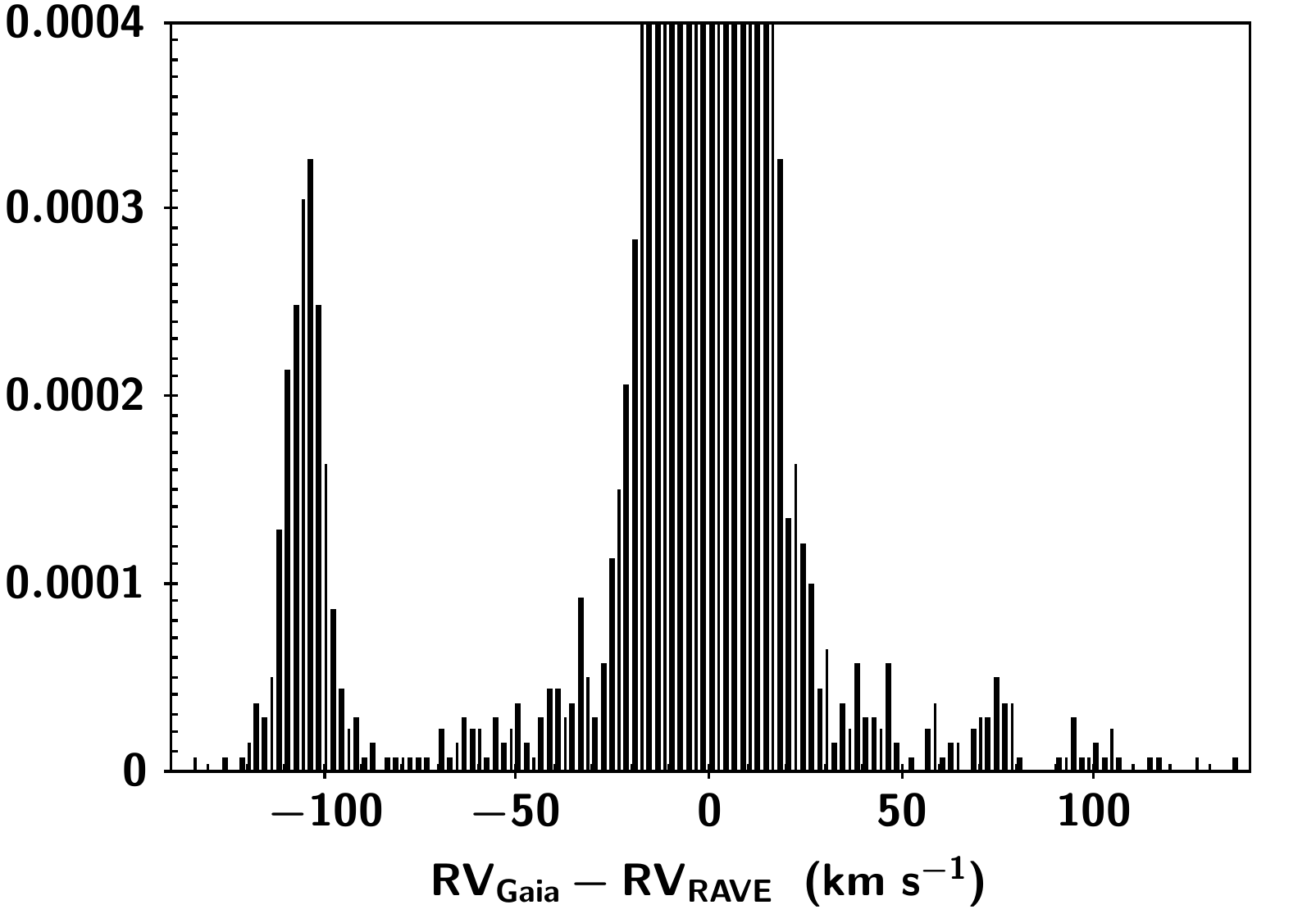}{0.25\textwidth}{(\emph{d})}
          }        
\caption{Distribution of $\rm RV_{RAVE}$ and $\rm RV_{Gaia}$ velocities for RAVE DR5 and Gaia DR2 cross-matched data with
(\emph{a}) no error limit,
(\emph{b}) SD$(\rm RV_{RAVE}) \leq$ 2 km s$^{-1}$, MAD$(\rm RV_{RAVE}) \leq$ 2 km s$^{-1}$,
(\emph{c}) SD$(\rm RV_{RAVE}) \leq$ 2 km s$^{-1}$, MAD$(\rm RV_{RAVE}) \leq$ 2 km s$^{-1}$, and cross-correlation error ($\sigma_{\rm RV}$) $\leq$ 2 km s$^{-1}$ or 2\% in $\rm RV_{Gaia}$ and $\rm RV_{RAVE}$.
(\emph{d}) A section of normalized histogram of difference in $\rm RV_{Gaia}$ and $\rm RV_{RAVE}$. Histogram is normalized with respect to maximum count and bin size used is 2 km s$^{-1}$.
\label{fig:Fig1}
}
\end{figure}

Radial velocity data of all the common stars between the two surveys is shown in Figure \ref{fig:Fig1}\emph{a}. Though most $\rm RV$ values 
agree well with each other, there are a number of stars for which differences between the two surveys are quite large, at the central portions in particular (see Figure \ref{fig:Fig1}\emph{a}).
Both the surveys provide $\rm RV$ along with the formal error, $\sigma_{\rm RV}$, which are measures of how well the cross-correlation of their spectrum is against the template spectrum. RAVE DR5 also provides standard deviation (SD) (its not same as $\sigma_{\rm RV}$) and the median absolute deviation (MAD), 
which provide independent measures of the $\rm RV_{RAVE}$ uncertainties calculated by re-sampling  a spectrum ten times.
For about 2.5\% of the RAVE sample, the difference in the radial velocity and radial velocity dispersions when spectrum is re-sampled 10 or 100 times is more than one-sigma (for more details see  \cite{Kunder2017AJ....153...75K}). 
On checking we found that these stars are the reason for very large scatter at the central portions as shown in Figure \ref{fig:Fig1}\emph{a}.
Considering the  typical accuracy of $\rm RV_{RAVE}$ in RAVE survey, which is better than 2 km s$^{-1}$ \citep{Kunder2017AJ....153...75K}, we excluded all those stars from the sample for which SD$(\rm RV_{RAVE}) >$ 2 km s$^{-1}$ and MAD$(\rm RV_{RAVE}) >$ 2 km s$^{-1}$. This resulted a total of 448,726 stars from RAVE DR5. 
The resultant data set of radial velocities from RAVE DR5 is compared in Figure \ref{fig:Fig1}\emph{b} with corresponding values of $\rm RV_{Gaia}$ from Gaia. 
Larger scatter that is present in Figure \ref{fig:Fig1}\emph{a} is almost absent in Figure \ref{fig:Fig1}\emph{b}. But one can notice a sharp  line parallel to the main line of  majority of stars for which $\rm RV$ values in both the catalogs are in good agreement. 

To further cull out the data with relatively large uncertainties (see Figure \ref{fig:Fig1}\emph{b}) and to retain good quality radial velocities, 
we chose only those stars from  both the surveys that have  $\sigma_{\rm RV}$ values either maximum of 2 km s$^{-1}$ (for retaining stars with small ${\rm RV}$)
or maximum percentage error of 2\% 
(for retaining stars with larger ${\rm RV}$)
from both the catalogs.
These filters yielded a sample of 322,878 stars.
Figure \ref{fig:Fig1}\emph{c} shows distribution of $\rm RV_{Gaia}$ and $\rm RV_{RAVE}$ velocities. In this figure, the parallel line substructure along with main line (where majority stars are lying) is clearly visible. The distribution of differences in $\rm RV_{Gaia}$ and $\rm RV_{RAVE}$ is shown as normalized histogram (Figure \ref{fig:Fig1}\emph{d}). 

\begin{figure}[ht!]
\includegraphics[width=0.5\textwidth]{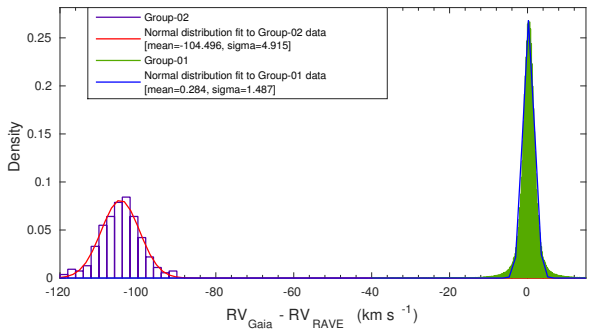}
\caption{Normalized distribution and corresponding normal distribution fit of difference in velocities for Group-01 and Group-02 stars. Histogram bin size for Group-01 and Group-02 are $0.2$ km s$^{-1}$ and $2$ km s$^{-1}$ respectively, and are in agreement with Freedman-Diaconis rule.}\label{fig:Fig2}
\end{figure}

Based on Figure \ref{fig:Fig1}\emph{c} and Figure \ref{fig:Fig1}\emph{d}, we divide the entire sample into three groups: (\emph{1}) {\bf{Group-01:}} the majority group consisting of 322,449 stars with a mean difference ($\rm RV_{Gaia} - \rm RV_{RAVE}$) of
$0.28$ km s$^{-1}$
with a spread given by $\sigma$= $1.49$ km s$^{-1}$ (Figure \ref{fig:Fig2}),
(\emph{2}) {\bf{Group-02:}} a small group consisting of about 272 stars with a large offset between $\rm RV_{Gaia}$ and $\rm RV_{RAVE}$, the normal distribution for the difference in $\rm RV_{Gaia}$ and $\rm RV_{RAVE}$ velocities is given in Figure \ref{fig:Fig2}, which shows that the mean difference between the velocities is $-104.50$ km s$^{-1}$ with spread given by $\sigma= 4.92$ km s$^{-1}$, and (\emph{3}) {\bf{Group-03:}}
the remaining 156 stars which lie on either side of the main group distribution.\\

In general, one would expect distribution in differences in high quality radial velocities between the two surveys similar to the main group (Group-01). However,  the large offset of $-$104 km $s^{-1}$ between the two surveys for the small group (group-02) is surprising.
There is a possibility of  mistakenly matching fore or background stars. For making sure that the $V_{rad}$ values of sample under consideration are in fact of the same stars in the respective surveys, we compared star's magnitudes and positions between the two surveys. To  eliminate this possibility, we make sure that the apparent magnitudes of common stars are same or within reasonable limits. But the problem is, the photometric pass-bands used for apparent magnitude measurements are not common in the two surveys.
Gaia DR2 provides photometric G, G$_{\rm BP}$, and G$_{\rm RP}$-band magnitudes in wavelength bands of [330-1050] nm, [330-680] nm, and [630-1050] nm respectively (see \cite{JordiGaiaPhotometry2010A&A...523A..48J}). RAVE does not have its own measurements, but it has magnitudes collected from various surveys such as Hipparcos, TYCHO2 and APASS. Of all, the AAVSO Photometric All-Sky Survey (APASS)
Data Release 9 (DR9) is the most comprehensive and precise with two Johnson broad band filters (B and V) and three Sloan filters. Survey is complete from 7 to 17th V-magnitude (hereafter $V_{\rm j,APASS}$) (see \cite{HendenApassLatestDataRelease2015AAS...22533616H}).
%
Cross-match between APASS and RAVE has been discussed in \cite{MunariRave2014AJ....148...81M}.
To compare $V_{\rm j,APASS}$ with Gaia's magnitude, both of these need to be put on the same scale. We converted Gaia's G-band photometric magnitudes to Johnson V magnitude (hereafter $V_{\rm j,Gaia}$) using conversion formulas provided by Gaia Collaboration (see \cite{EvansGaiaDr2Photo2018arXiv180409368E}).

\begin{figure}[ht]
\gridline{
			\fig{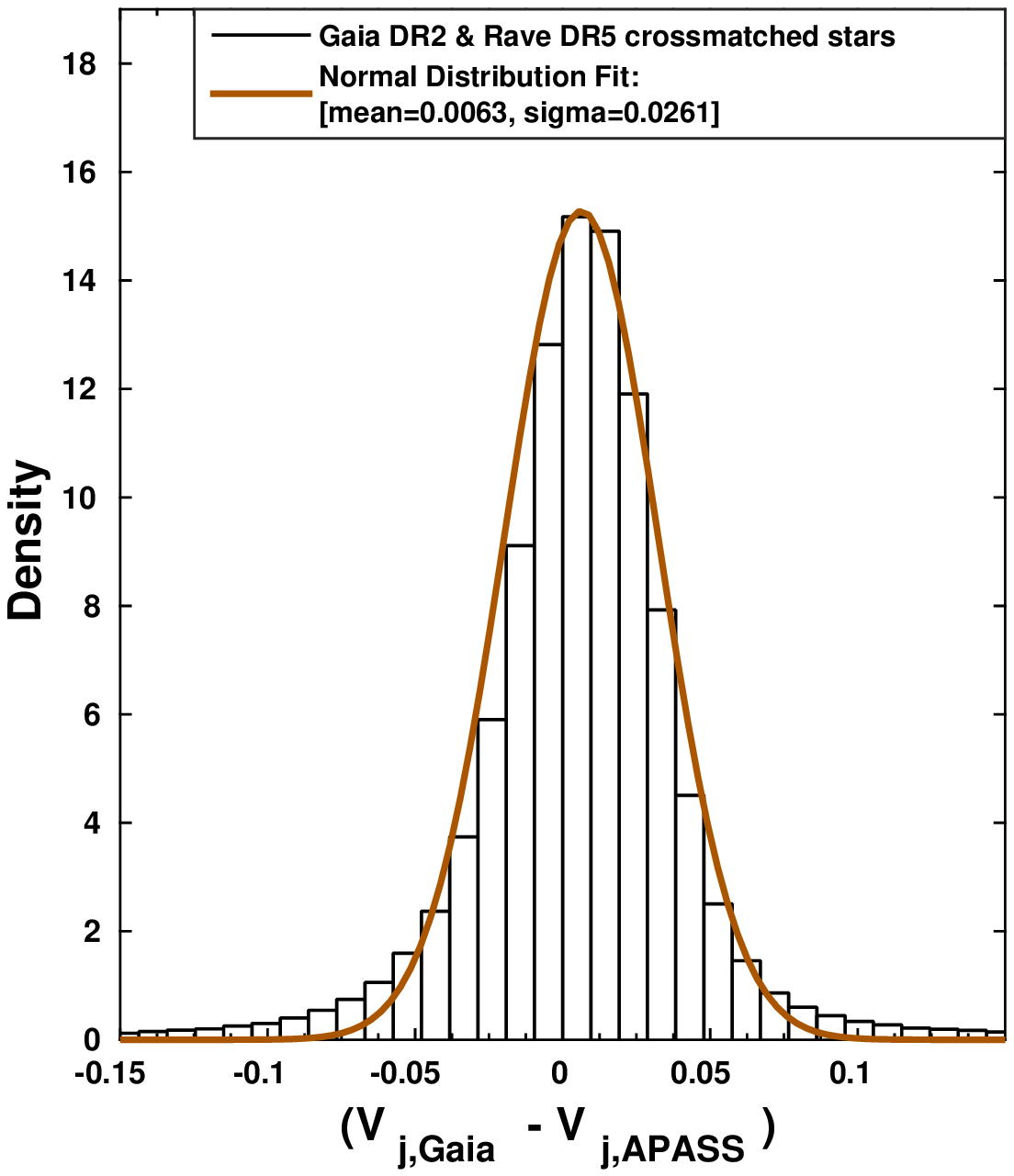}{0.25\textwidth}{(\emph{a})} 
			\fig{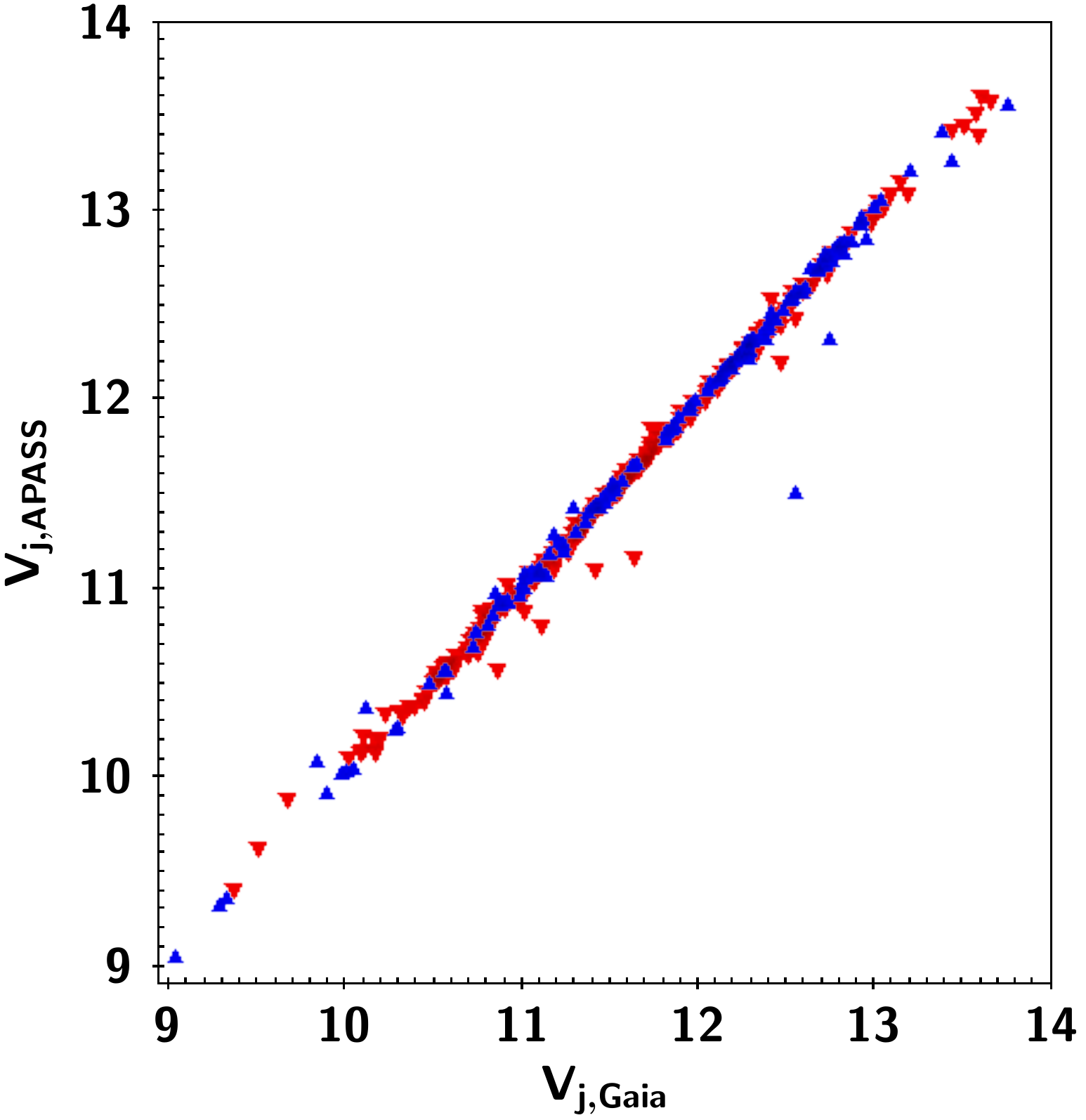}{0.25\textwidth}{(\emph{b})}
}
\caption{
(\emph{a}) Distribution of difference between $V_{\rm j,Gaia}$ and $V_{\rm j,APASS}$ for all the Gaia DR2 and RAVE DR5 cross-matched stars for which both $V_{\rm j,Gaia}$ and $V_{\rm j,APASS}$ are available (441,106 stars), and corresponding normal distribution fit.
(\emph{b}) Distribution of $V_{\rm j,APASS}$ and $V_{\rm j,Gaia}$ for stars belonging to Group-02 (red) and Group-03 (blue). 
Here, $V_{\rm j,Gaia}$ is calculated from Gaia's G-band magnitudes and $V_{\rm j,APASS}$ is provided in RAVE DR5.
\label{fig:Fig3}
}
\end{figure}

\begin{figure}[ht!]
\gridline{
			\fig{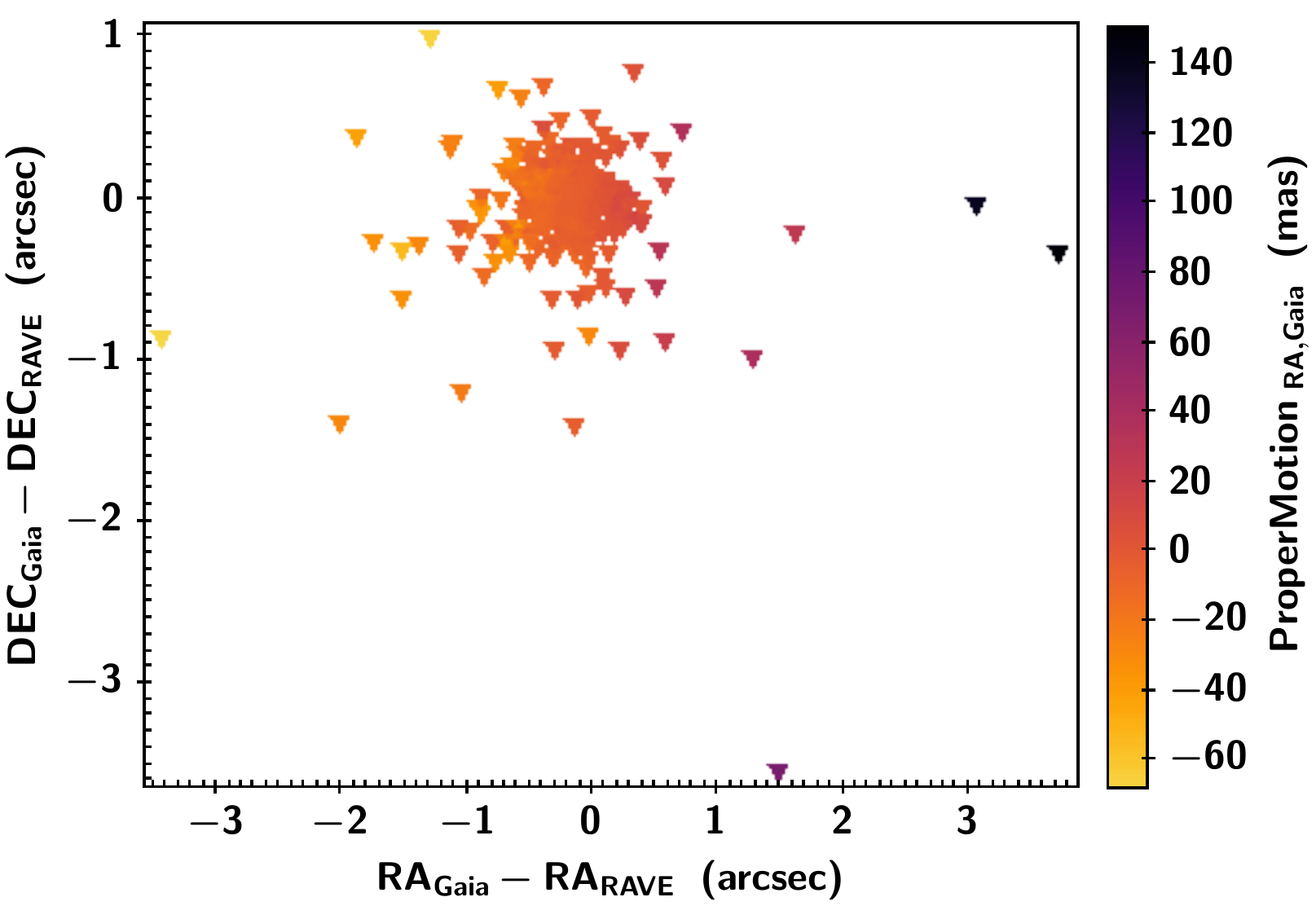}{0.25\textwidth}{(\emph{a})}
			\fig{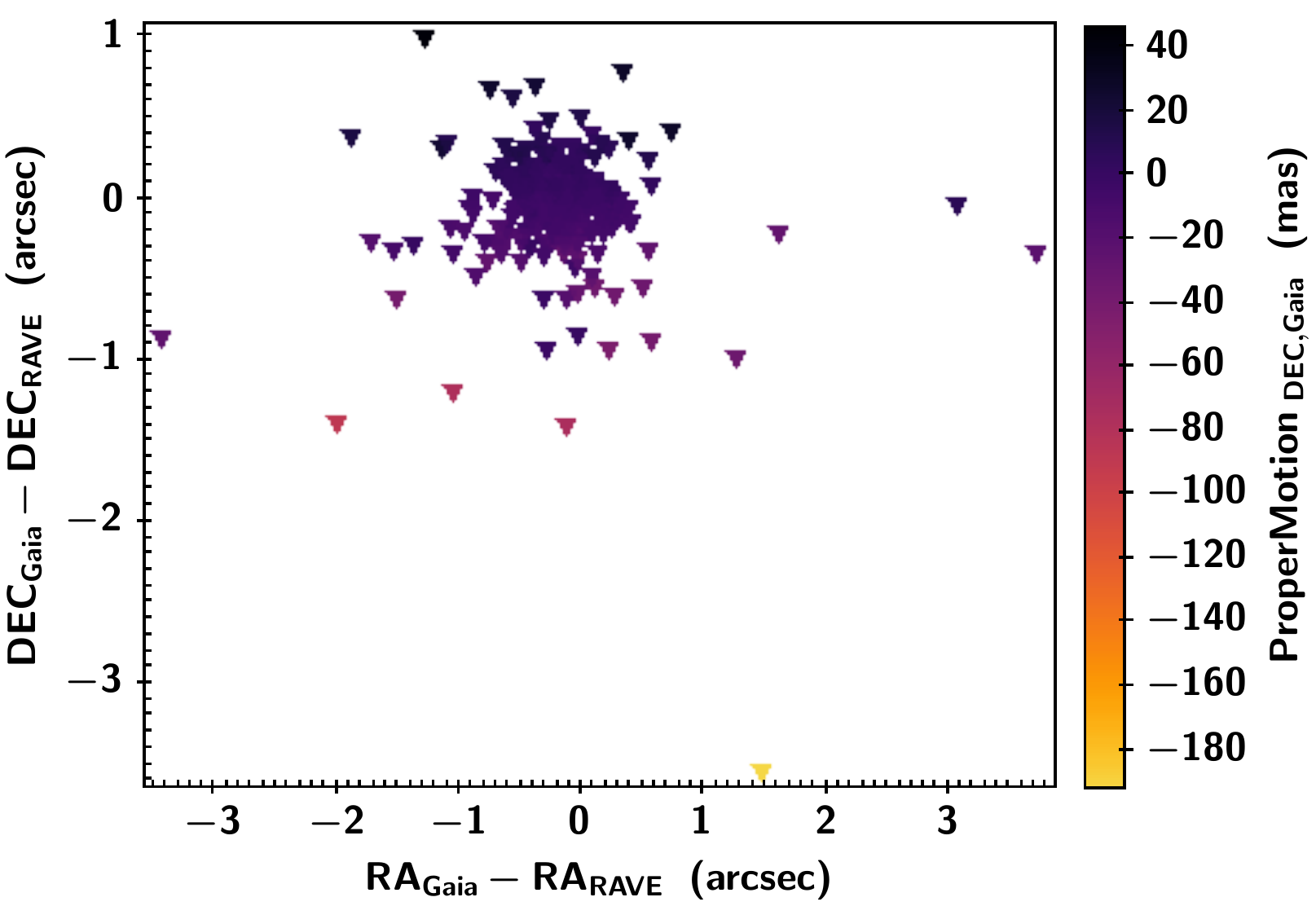}{0.25\textwidth}{(\emph{b})}
          }
\gridline{
			\fig{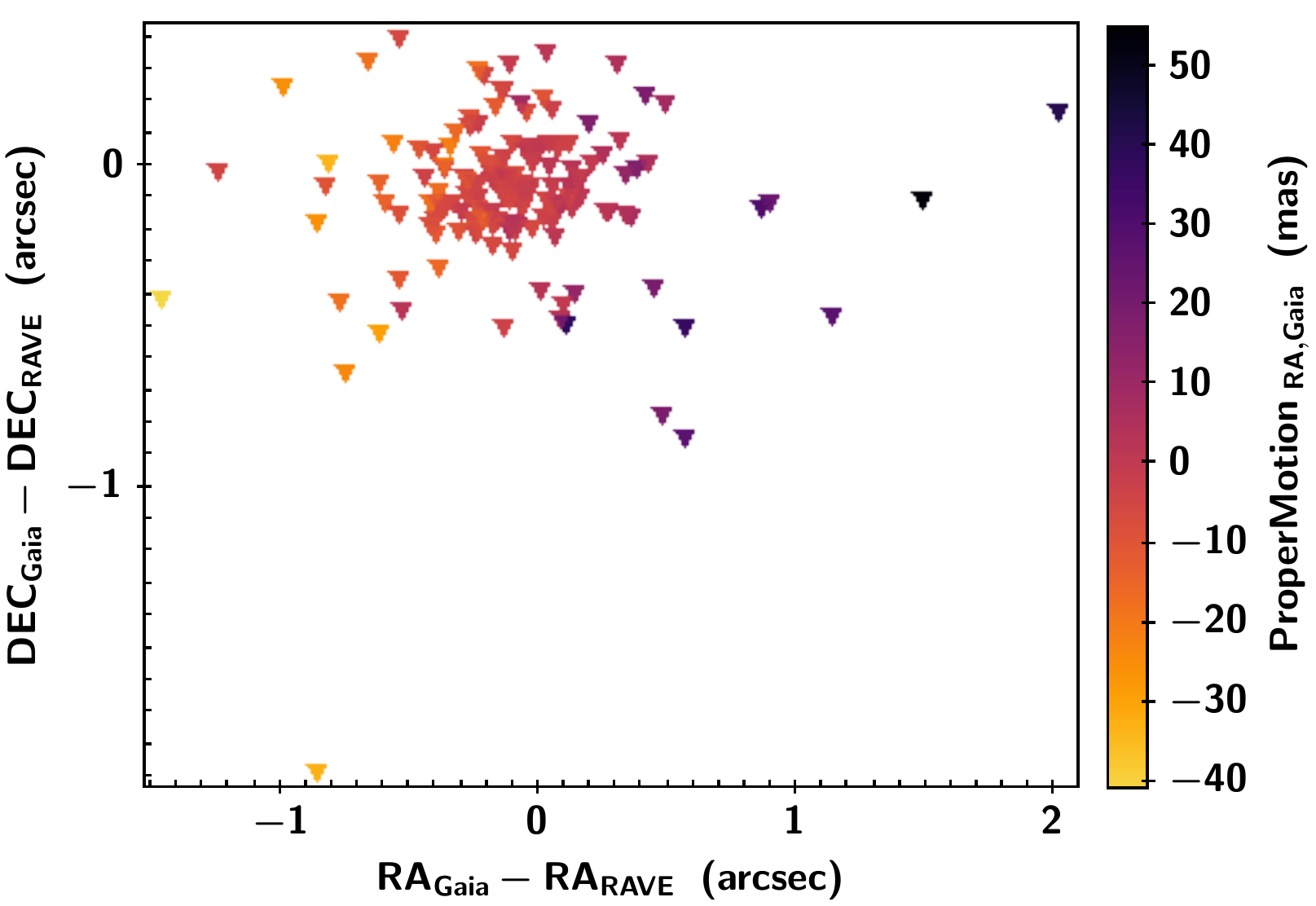}{0.25\textwidth}{(\emph{c})}
			\fig{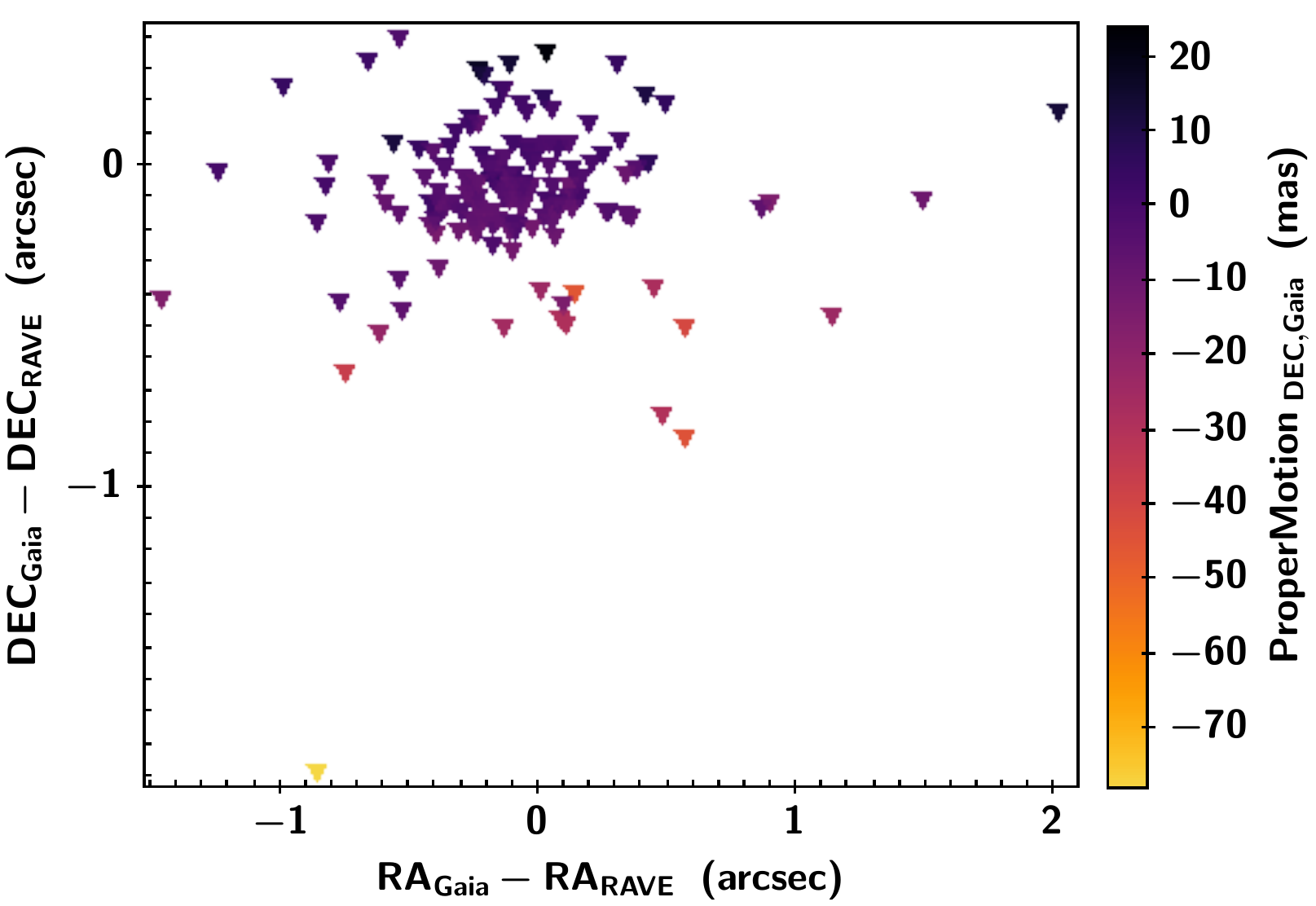}{0.25\textwidth}{(\emph{d})}
          }
\caption{Distribution of difference in position coordinates of Group-02 stars from Gaia DR2 and RAVE DR5 catalog when corresponding proper motions (pm) from Gaia DR2 along
(\emph{a}) right ascension (RA), and 
(\emph{b}) declination (DEC) are used as weights.
Distribution of difference in position coordinates of Group-03 stars from Gaia DR2 and RAVE DR5 catalog when corresponding proper motions (pm) from Gaia DR2 along
(\emph{c}) right ascension (RA), and 
(\emph{d}) declination (DEC) are used as weights.
\label{fig:Fig4}
}
\end{figure}

In Figure \ref{fig:Fig3}, we plotted the distribution of magnitude difference ($V_{\rm j,Gaia} - V_{\rm j,APASS}$) for the entire sample of stars for which magnitudes are available in both the surveys. Distribution shows that the magnitudes in both the surveys agree well with a mean difference of 0.006 with sigma = 0.026 (see Figure \ref{fig:Fig3}\emph{a}). Magnitude agreement between the two data sets with the exemption of a couple of outliers (Figure \ref{fig:Fig3}\emph{b}),  validates cross-matching of stars in the two surveys provided by RAVE.
This shows that the systematic difference in $\rm RV$ is not an artifact arising from mismatch of stars between the two surveys. For stars of Group-01 and Group-02, comparison of positions of stars (RA $\&$ DEC) in the surveys are given in Figure \ref{fig:Fig4}\emph{a}, \ref{fig:Fig4}\emph{b} and Figure \ref{fig:Fig4}\emph{c}, \ref{fig:Fig4}\emph{d} respectively. Difference in position is quite small, except in cases where stars have comparatively large proper motion. Given the different epochs of surveys, such small differences in position are expected.  

Of 272 stars in Group2, there are
51 stars for which radial velocities are given in RAVE DR5 from more than one spectrum observed at different times. Out of these 51 stars with multiple spectras, on checking $\rm RV$ for the same star from different RAVE observations, we found that 50 stars have at least one spectrum which gave almost similar values as $\rm RV_{Gaia}$ along with the ones which gave a difference of approximately $-$104km/sec. Exception is the star `Gaia DR2 5913047541322494080' for which the two listed velocities in RAVE DR5 from two different spectras (`20100803\_1726m56\_003' and `20100731\_1726m56\_003') differ from that of $\rm RV_{ Gaia}$ by approximately $-$107 km s$^{-1}$. Our understanding is that the same reduction methodology was used to derive the radial velocities from the multiple spectra for the same star, and yet for a small group there seems to be a problem.

\section{Discussion}
The relatively small spread in radial velocities  in case of the main group (i.e. Group-01) is probably due to intrinsic errors related to different instrumental set-ups and measurement methods.
One also cannot rule out the possibility of such  differences as a result of measuring intrinsic random motions of stars including low-amplitude pulsations
(both radial and non radial) frequently present on giant stars, the spectroscopic binaries and gravitational redshifts at two different epochs \citep{LindegrenRvFundamentals2003A&A...401.1185L}.
The Gaia DR2  data provides median radial velocities averaged over first 22 months of observations since its launch in July 2014.
On the other hand, RAVE velocities are from the spectra observed from April 2003 to April 2013. The two surveys measured radial velocities in a time difference of about 2 to 12 years.\\

However, it is not clear why such large offset  exists for stars of Group-02 between the two surveys. 
Though the fraction of faulty stars forms a very small percentage ($\approx 0.08\%$) of total sample of highly accurate radial velocity stars considered,
it is important to highlight the issue of discrepancy to avoid misleading results. For example, Kinematic velocities (U, V, W) computed using $\rm RV$ values from RAVE DR5 and Gaia DR2 for Group-02 stars differ significantly.
\begin{figure}[ht!]
            \fig{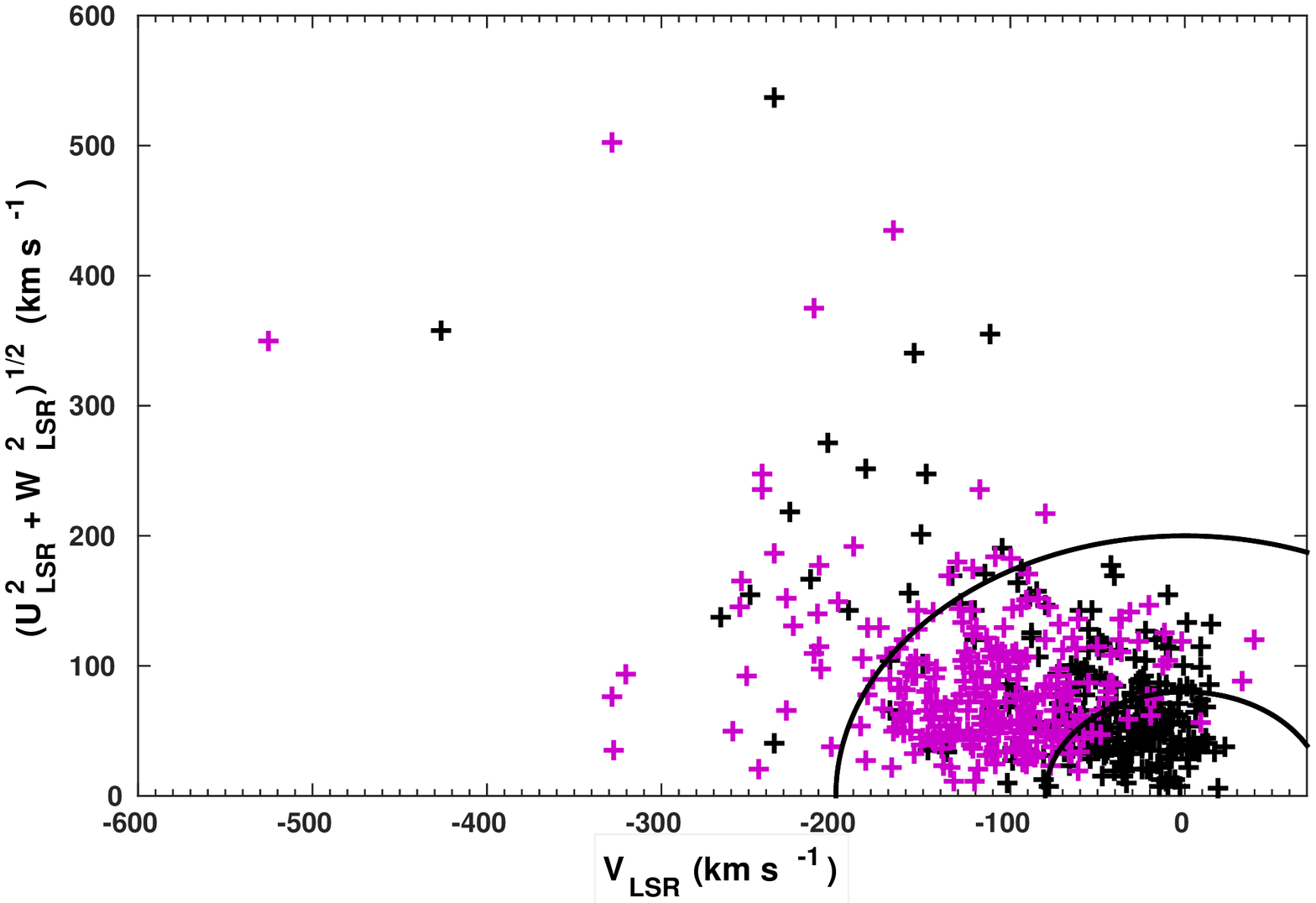}{0.5\textwidth}{(\emph{a})}
          	\fig{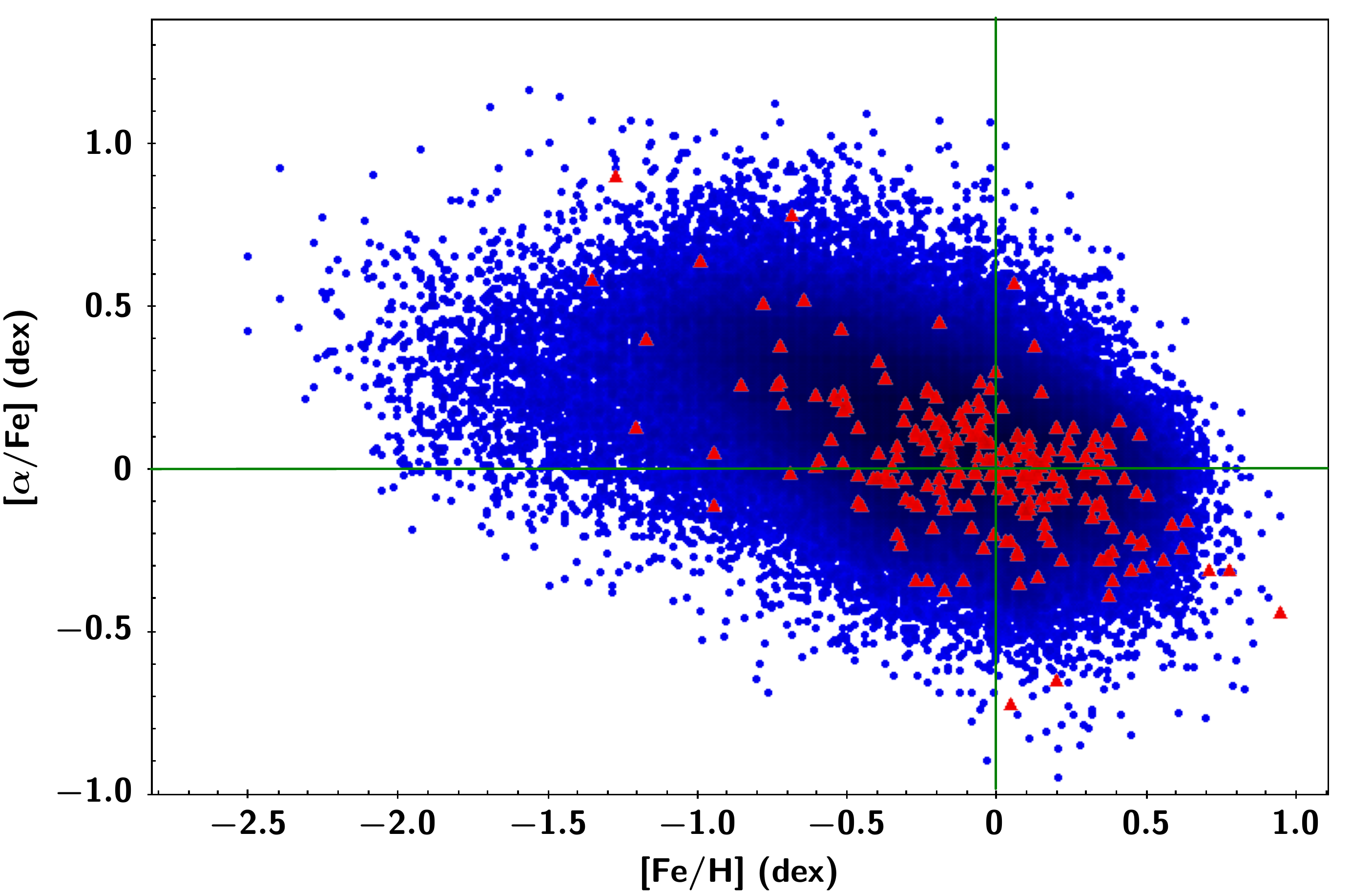}{0.47\textwidth}{(\emph{b})}
\caption{
(\emph{a}) Toomre diagram for stars in Group-02 with kinematics deduced using $\rm RV_{Gaia}$ (black) and $\rm RV_{RAVE}$ (magenta). The two concentric circles delineate constant peculiar velocities (V$_{\rm tot}$ = (U$^{2}_{\rm LSR}$ + V$^{2}_{\rm LSR}$ + W$^{2}_{\rm LSR}$)$^{1/2}$) of 80 km s$^{-1}$ and 200 km s$^{-1}$, and represent kinematic boundary for thin disk ($|$V$_{\rm tot}| <$ 80 km s$^{-1}$), thick disk (80 $<|$V$_{\rm tot}| <$ 200 (km s$^{-1}$)) and halo ($|$V$_{\rm tot}| >$ 200 (km s$^{-1}$)) which are in accordance with results in \cite{Reddy2006MNRAS.367.1329R}. 
(\emph{b}) Distribution of [$\alpha$/Fe] with respect to [Fe/H] of Group-02 stars (red) with Group-01 in background (blue).
Measurements of [$\alpha$/Fe] are given in RAVE DR5, and out of 272 stars of Group-02, both [$\alpha$/Fe] and [Fe/H] are available for 227 stars which are plotted here.
\label{fig:Fig5}
}
\end{figure}
Probabilities computed based on the two different sets of kinematics using the recipe given in \cite{Reddy2006MNRAS.367.1329R} lead to different Galactic components to which stars belong. Stars kinematics based on $\rm RV_{ Rave}$ suggest that most of the stars belong to the thick disk. On the other hand,  the kinematics based on $\rm RV_{ Gaia}$ values suggest that most of the stars belong to the Galactic thin  disk component. This has been illustrated in Figure \ref{fig:Fig5}\emph{a} in the form of Toomre diagram which represents the relationship between the sum in quadrature of the vertical and radial velocities (i.e. kinetic energy) and the rotational velocity (i.e. rotational energy) relative to the local standard of rest \citep{SandageToomreDiagram1987AJ.....93...74S}.
Star's heliocentric velocities (U, V, W) are corrected for the solar motion (using U$_{o}$=10, V$_{o}$=5.3, W$_{o}$=7.2 (km s$^{-1}$) from \cite{DehnenBinneyLSR1998MNRAS.298..387D}) to get velocities with respect to local standard of rest (U$_{\rm LSR}$, V$_{\rm LSR}$, W$_{\rm LSR}$).
Also, the  used kinematic boundaries for thin disk ($|$V$_{\rm tot}| <$ 80 km s$^{-1}$), thick disk (80 $<|$V$_{\rm tot}| <$ 200 (km s$^{-1}$)) and halo ($|$V$_{\rm tot}| >$ 200 (km s$^{-1}$)) are in accordance with the results in \cite{Reddy2006MNRAS.367.1329R}.  

To understand the source for this discrepancy, we examined star's metallicity ([Fe/H]) provided by RAVE DR5. The distribution shows that the [Fe/H] ranges from $-$1.2 to $+$0.5 dex (Figure \ref{fig:Fig5}\emph{b}) which is typical of the Galactic disk metallicity range \citep{Reddy2006MNRAS.367.1329R}. Combined with the available [Fe/H] and kinematics implies that the stars  may belong to the Galactic thin disk and the kinematics based on velocities from Gaia seems to be consistent with the star's [Fe/H] distribution. Though thick disk [Fe/H] overlaps with the thin disk metallicities, one would expect most of the stars in the metal-poor side beyond $-$1.2~dex in case of thick disk. Another evidence could be $\alpha$-process elements for distinguishing thin disk stars from thick disk \citep{Reddy2003MNRAS.340..304R}. Abundances taken from RAVE survey are plotted against [Fe/H] (Figure \ref{fig:Fig5}). 
Though majority of stars do show normal $\alpha$-process, it does not provide clear separation. 

\startlongtable
\begin{table*}[ht!]
\begin{deluxetable}{|ccc|ccc|ccc|}
\tablecaption{$\rm RV$ of three common stars from Gaia, RAVE and LAMOST of Group-02.\label{tab:table1}}
\tablehead{
 & \colhead{Object ID} &  &  &\colhead{$\rm RV$} (km s$^{-1}$)& & &\colhead{$\sigma_{\rm RV}$} (km s$^{-1}$)&\\
Gaia DR2 & RAVE & LAMOST & Gaia & RAVE & LAMOST & Gaia & RAVE & LAMOST
}
\startdata
5763571271979913472   &   20060325\_0853m01\_007   &   309408135   &   60.47   &   164.54   &   56.7   &   0.49   &   0.78   &   -   \\
3831262427493102976   &   20110422\_1013m00\_108   &   230002091   &   79.43   &   196.79   &   84.5   &   1.05   &   1.57   &   -  \\
3831274998862303360   &   20110422\_1013m00\_107   &   310212200   &   21.66   &   132.48   &   19.1   &   1.63   &   1.48   &   -  \\
\enddata
\end{deluxetable}
\end{table*}

Further we searched for radial velocity measurements of Group-02 stars among LAMOST and APOGEE spectroscopic surveys. Unfortunately, we found just three stars that are common with LAMOST survey. In Table \ref{tab:table1}, we have summarized $\rm RV$ values of these three common stars from all the three surveys. LAMOST values match well with those from the Gaia data suggesting that the large offset is probably caused by RAVE data set.

\begin{figure}[ht]
			\fig{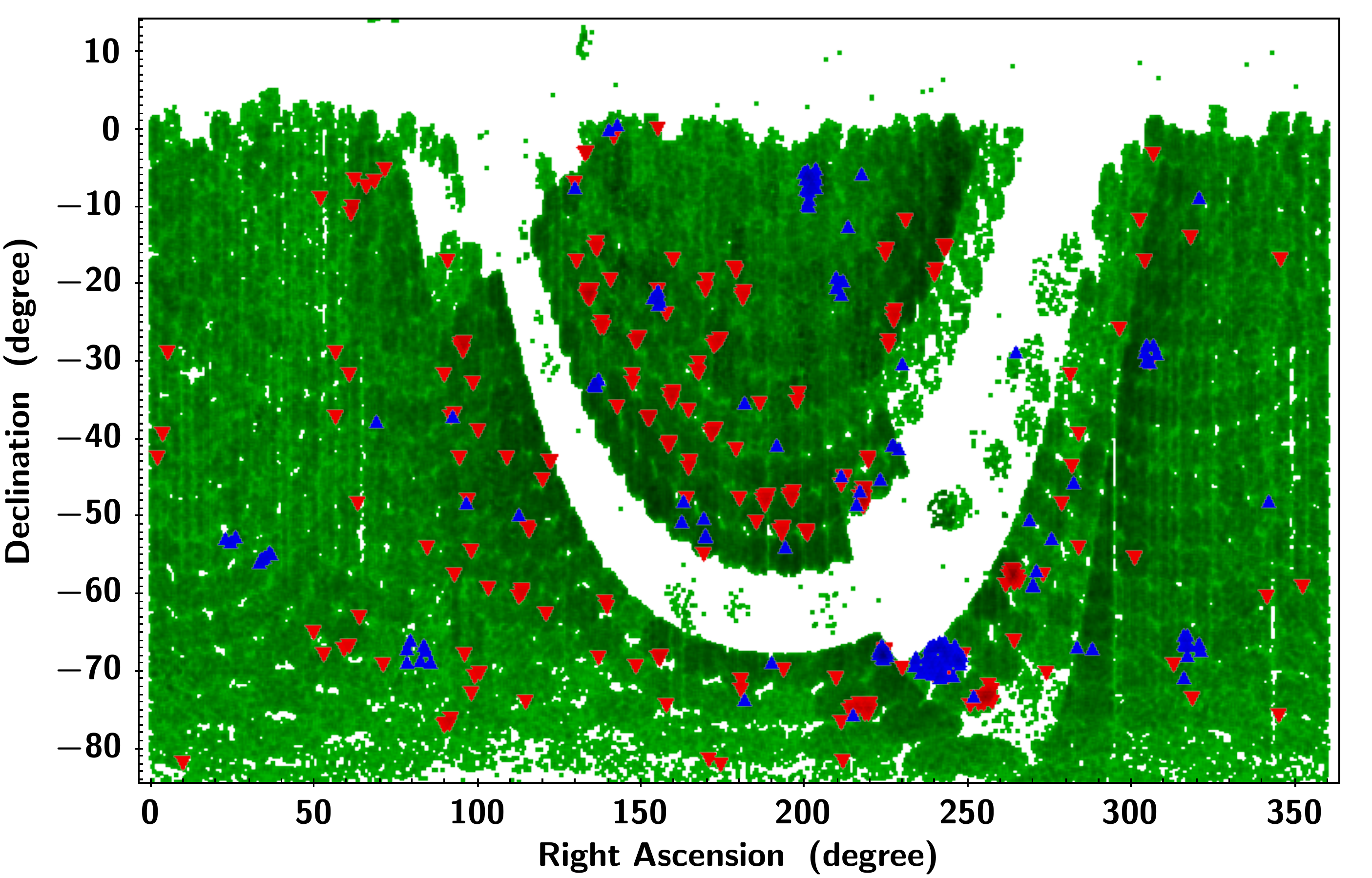}{0.485\textwidth}{(\emph{a})}
          	\fig{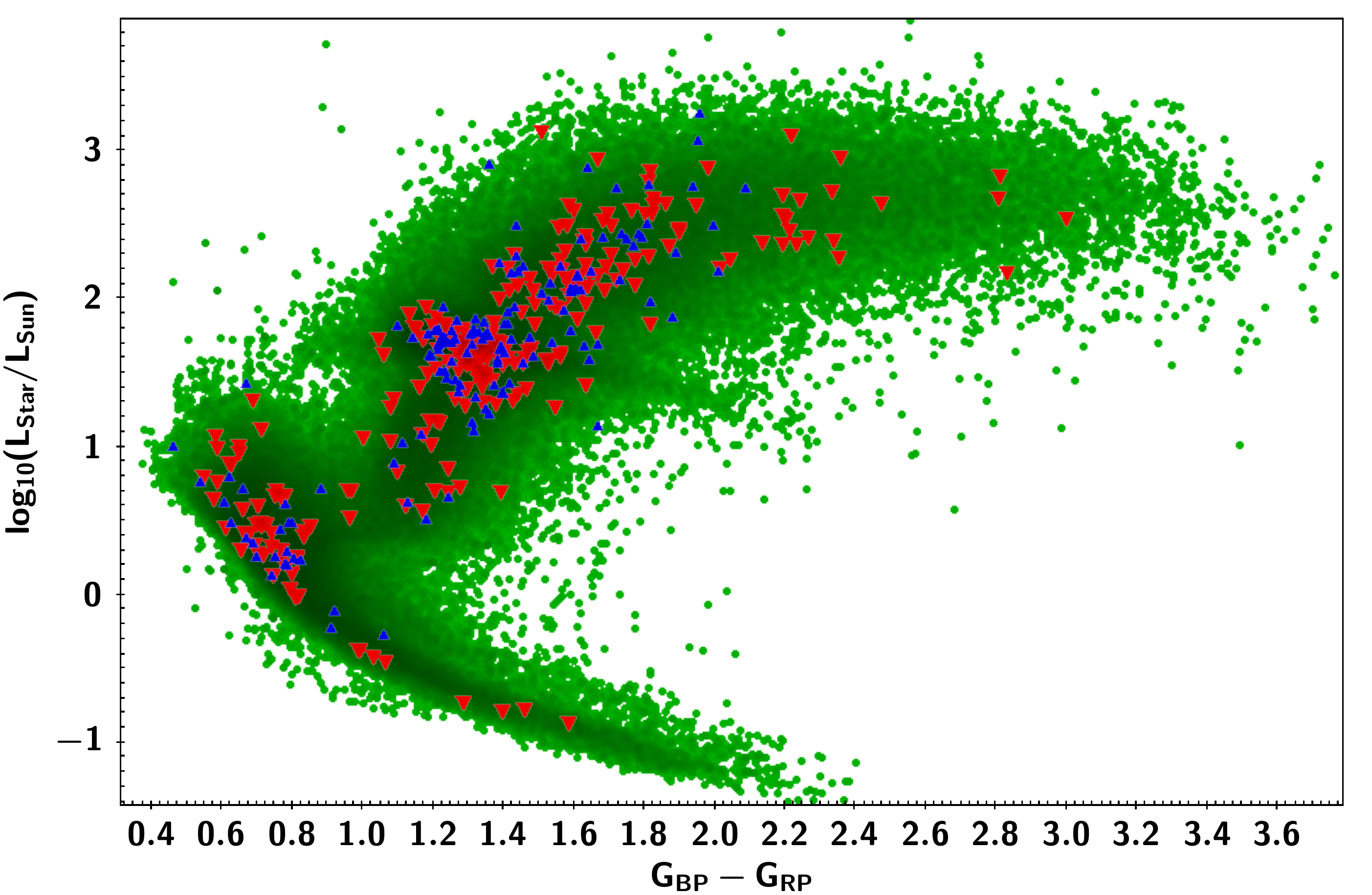}{0.48\textwidth}{(\emph{b})}
\caption{(\emph{a}) Sky distribution, and (\emph{b}) Color-Luminosity diagram for stars in our sample. Here green, red and blue color represent Group-01, Group-02 and Group-03 stars respectively.
\label{fig:Fig6}
}
\end{figure}

The more intriguing part is the near constant offset of $-$104.5  km s$^{-1}$ with a small dispersion. To search for clues for the discrepancy in radial velocity data, we examined whether these stars were of some particular kind or localized in space.  In Figure \ref{fig:Fig6}\emph{a}, we showed Group-02 $\&$ 03 stars superposed with the entire sample as background. Stars are all over the sky without any spatial clustering ruling out the possibility that these stars belong to spatially localized cluster or clusters. We also checked whether these stars are of any particular type. As shown in Figure \ref{fig:Fig6}\emph{b}, distribution of stars in HR-diagram suggest that they  are uniformly distributed across the stellar evolutionary phases and have no particular trend with either T$_{\rm eff}$ or log $g$ with respect to the main group.

Another possibility may be the cluster evaporation, in particular the short lived open cluster with loosely bound member stars which may become disrupted and the member stars escape  due to encounters with other massive structures in the galaxy like clusters and clouds of gas, and tidal force in the galactic gravitational field as they orbit the galactic center (\cite{TrumplerOpenCluster1930LicOB..14..154T}, \cite{ShapleyStarCluster1930HarMo...2.....S}, \cite{WielenStarCluster1988IAUS..126..393W}).
Members of such open clusters will be spread along the path traced by it and continue to orbit the galaxy  with inherited velocities with certain dispersion \citep{TrumplerOpenCluster1930LicOB..14..154T}.
However, this will not explain the offset in difference between the two surveys, but will provide a clue that the members might have belonged to a particular group in the past.
A closer scrutiny of data reveals that these stars are
at different distances ranging from 80 pc to 3 Kpc (approximately) and also they don't seem to be kinematically similar as well.  Thus, it is unlikely that the offset in difference is due to the stars that belonged to a single group either in the past or present. 

\section{conclusion}
While comparing radial velocity data from the two major surveys RAVE and Gaia, we noticed a significant difference in radial velocity with an offset of $-$104.5 km s$^{-1}$ for a small group of 272 stars. While kinematics based on RAVE suggest that most of the stars in the group are of thick disk, velocities from Gaia suggest the stars are of thin disk. However, [Fe/H] range of stars in the group from $-$1.2 to +0.5 dex suggests that most of the stars are in fact from thin disk origin agreeing with Gaia velocity data. This is corroborated by LAMOST velocity data for three common stars, in all three surveys, which are in agreement with Gaia values.
Though source for the offset is not clear, our study suggests that the offset is due to RAVE data set.
\footnote{In refereeing we became aware that the RAVE Consortium independent of our analysis has released a list with affected targets on July 13, 2018 (available at \url{https://www.rave-survey.org/project/documentation/dr5/rave_dr5/}), and gave issue with a faulty wavelength calibration as the reason for the offset. A corrected wavelength calibration and derived radial velocity will be available for affected targets with the soon-to-be released final RAVE DR6.}     
 
\section{Acknowledgments} \label{sec:acknowledgments}
{\footnotesize
We thank Dr. Giovanni Carraro (AAS scientific editor) and anonymous reviewers for their constructive comments and suggestions, which helped us to improve the manuscript. 
This work has made use of data from the European Space Agency (ESA) mission
{\it Gaia} (\url{https://www.cosmos.esa.int/gaia}), processed by the {\it Gaia}
Data Processing and Analysis Consortium (DPAC,
\url{https://www.cosmos.esa.int/web/gaia/dpac/consortium}). Funding for the DPAC
has been provided by national institutions, in particular the institutions
participating in the {\it Gaia} Multilateral Agreement.
}
{\footnotesize
This work has also made use of RAVE data. Funding for RAVE (www.rave-survey.org) has been provided by institutions of the RAVE participants and by their national funding agencies.
}

\bibliographystyle{aasjournal}
\bibliography{ref}

\end{document}